\documentclass[]{raa}

\usepackage{graphicx,times}
\usepackage{natbib}
\usepackage{amssymb,amsmath}
\bibpunct{(}{)}{;}{a}{}{,}

\usepackage[pagebackref=true]{hyperref}

\begin{document}
  
    \title{Basic Survey Scheduling for the Wide Field Survey Telescope (WFST)}

  \volnopage{ {\bf 2023} Vol.\ {\bf X} No. {\bf XX}, 000--000}
    \setcounter{page}{1}

   \author{Yan-Peng~Chen\inst{1,2}, Ji-an~Jiang\inst{3,4}, Wen-Tao Luo\inst{5,3}, Xian~Zhong~Zheng\inst{1,2,}\thanks{Corresponding author.}, Min~Fang\inst{1,2}, Chao~Yang\inst{1,2}, Yuan-Yu~Hong\inst{1,2}, Zong-Fei~L\"{u}\inst{1,2}
   }
%% Here is an example of three authors come from different institutes.
%% For single author or all the authors from an institute, use "\inst{}" only

\institute{ Purple Mountain Observatory, Chinese Academy of Sciences, Nanjing 210023, China {\it xzzheng@pmo.ac.cn}\\
\and School of Astronomy and Space Science, University of Science and Technology of China, Hefei 230026, China \\
\and Department of Astronomy, University of Science and Technology of China, Hefei 230026, China \\
\and National Astronomical Observatory of Japan, National Institutes of Natural Sciences, 2-21-1 Osawa, Mitaka, Tokyo 181-8588, Japan \\
\and Institute of Deep Space Sciences, Deep Space Exploration Laboratory, Hefei, 230026, China\\
%% Please give the E-mail address of the author, to whom future correspondence and
%% offprint requests will be sent.
  %       \and
  %            Yunnan Astronomical Observatory, National Astronomical Observatories, Chinese Academy of Sciences,
  %            Kunmin 650011, China\\
	% \and
%	  Center for Astrophysics, University of Science and Technology of China, Hefei 230026, China\\
% Key Laboratory for Research in Galaxies and Cosmology, The University of Science
% and Technology of China, Chinese Academy of Sciences, Hefei, Anhui, 230026, China\\
% \and 
% Polar Research Institute of China,
% Jinqiao Rd. 451, Shanghai, 200136, China\\
\vs \no
    {\small Received 2023 Month Day; accepted 202X Month Day}
}

\abstract{
Aiming at improving the survey efficiency of the Wide Field Survey Telescope, we have developed a basic scheduling strategy that takes into account the telescope characteristics, observing conditions, and weather conditions at the Lenghu site. The sky area is divided into rectangular regions, referred to as `tiles', with a size of $2.577\degr \times 2.634\degr$ slightly smaller than the focal area of the mosaic CCDs. These tiles are continuously filled in annulars parallel to the equator. The brightness of the sky background, which varies with the moon phase and distance from the moon, plays a significant role in determining the accessible survey fields. Approximately 50 connected tiles are grouped into one block for observation. To optimize the survey schedule, we perform simulations by taking into account the length of exposures, data readout, telescope slewing, and all relevant observing conditions. We utilize the Greedy Algorithm for scheduling optimization. Additionally, we propose a dedicated dithering pattern to cover the gaps between CCDs and the four corners of the mosaic CCD array, which are located outside of the $3\degr$ field of view. This dithering pattern helps to achieve relatively uniform exposure maps for the final survey outputs.
\keywords{ techniques: scheduler — survey — telescope — method
}
}
  \authorrunning{Y.-P. Chen et al.}            %author_head in even pages
  \titlerunning{Basic Survey Scheduling of the Wide Field Survey Telescope (WFST)}  % title_head in odd pages
  \maketitle

%________________________________________________ sections below
% 

\twocolumn

\section{Introduction}           %% first-level sections will be auto-capitalized
\label{sect:1}

Survey scheduling is a crucial task in conducting large-scale astronomical surveys, particularly for all-sky imaging surveys with dedicated wide-field telescopes designed for multiple purposes. The Hyper Suprime-Cam Subaru Strategic Program (HSC-SSP) \citep{Aihara2018PASJ} was conducted using the 8.2m Subaru telescope and involved a three-layer (wide, deep, and ultra-deep) imaging survey. The primary goals of this survey were to explore the dark matter and dark energy of the Universe, study galaxy evolution, investigate the intergalactic medium, and observe transients at high redshifts. A dithering pattern was applied to ensure uniform exposure time for each tract on the observed sky map, with each tract representing the basic unit of one pointing.

On the other hand, the Zwicky Transient Facility (ZTF) \citep{bellm2018zwicky, bellm2019zwicky} focuses on detecting time-domain astronomical events such as supernovae, tidal disruption events (TDEs), active galactic nuclei (AGNs), and the electromagnetic counterparts of gravitational wave events. With its extremely large field of view (47\,deg$^2$), ZTF is capable of scanning the entire northern sky every three days, maximizing its ability to capture bright transient sources.

The Legacy Survey of Space and Time (LSST) \citep{ivezic2019lsst} of the Vera C. Rubin observatory aims to map a survey area of 20,000\,deg$^2$ in the southern hemisphere and achieve a detection limit of $r$=28.0\,mag over a period of ten years with 180 visits. This survey will offer unprecedented data not only for dark matter and dark energy research but also for the study of various transient phenomena. The survey strategies will be meticulously scrutinized based on metrics designed by different science working groups, including the basic science requirement document (SRD) metrics and the dark energy science collaboration (DESC) wide fast deep field (WFD) metric, which encompasses supernovae, TDEs, and fast microlensing.

The Wide Field Survey Telescope (WFST) is an imaging facility jointly operated by the University of Science and Technology of China (USTC) and the Purple Mountain Observatory (PMO), Chinese Academy of Sciences. Located at the Lenghu site in Qinghai Province, the telescope has a primary mirror diameter of 2.5 meters and an effective field of view of 6\,deg$^2$. Its primary scientific objectives include the study of extragalactic transients such as supernovae, TDEs, and AGNs. Additionally, the Target of Opportunities (ToO) time is dedicated to searching for electromagnetic/optical counterparts of gravitational wave events, gamma-ray bursts, and fast radio bursts whenever there is a high possibility of detection. Other key projects include the study of Near Earth Objects (NEOs), Milky Way satellites, galaxy formation, and cosmology. We have designed a basic survey strategy to maximize the instrument's capabilities and ensure optimal results, taking into consideration various factors (\citealp{wfst2023}; Jiang et al. 2023, in prep).

WFST will carry out a six-year observing program consisting of a wide-field imaging survey and a deep high-cadence survey in the $u$, $g$, $r$, $i$ and $z$ bands. These two key surveys will utilize approximately 90\% of the total observing time, with the remaining time allocated to specific survey programs. Before initiating any large-scale survey programs, it is crucial to determine the telescope's observation capabilities and develop appropriate survey scheduling that considers both scientific requirements and observing conditions throughout the night. We specifically focus on factors essential for imaging quality, such as site and meteorological conditions, moon phase and distance to target fields, as well as telescope and camera operation time (e.g., slew and readout time). Achieving these objectives relies on the implementation of sophisticated algorithms.

Neural network-based optimization algorithms have been employed in the observation scheduling of the Hubble Space Telescope \citep{johnston1992scheduling}. The RTS2 (Remote Telescope System 2nd Edition) system, an open source software for autonomous telescope observations, is well-known in the field. Initially, it implemented a simple value function for survey scheduling optimization but later improved to incorporate genetic algorithms \citep{kubanek2010genetic}. \citet{bellm2019zwicky} optimized the survey plan for ZTF using an integer linear programming algorithm, while \citet{naghib2019framework} utilized the Markov Decision Process to address the survey scheduling problem for the Vera C. Rubin Observatory (LSST; \citealt{ivezic2019lsst}).

In this paper, we present a basic survey scheduler for WFST based on the Greedy Algorithm. This scheduler can be iteratively refined as we gain a better understanding of the observing conditions and facility characteristics. Additionally, we demonstrate that employing a specific dithering pattern for survey observations can significantly improve the uniformity of imaging data. To evaluate the performance of the scheduler, we generate semi-realistic conditions by adjusting parameters to model various moon phases and atmospheric conditions at an altitude of 4,170 meters, representative of the WFST site. The structure of this paper is as follows: In Sect.~\ref{sect:2}, we introduce the simulation methodology used to replicate the WFST site and prepare for the design of survey scheduling and optimization. In Sect.~\ref{sect:3}, we present the scheduler based on the Greedy Algorithm. We introduce an optimized dithering pattern to mitigate data loss resulting from CCD gaps in Sect.~\ref{sect:4}. Finally, we provide a summary of our findings and outline future work in Sect.~\ref{sect:5}.

\section{Simulation Preparation for WFST Scheduler Design}
\label{sect:2}

Several factors must be taken into consideration when optimizing the scheduling scheme for WFST surveys. The observing conditions at the telescope site are of utmost importance, particularly the sky background under varying moon conditions (moon phase and distance), airmass, seeing, and clean night fraction. Additionally, the parameters of the telescope and the prime-focus camera play a critical role in scheduler design, including factors such as telescope movement and stabilization time (referred to as slew time), as well as the camera's readout and filter changing time. However, other conditions, such as humidity, cloud coverage and other weather factors are not included here, which will be monitored and handled by the Observatory Control System (OCS).

Here, we present the key observational conditions specific to the Lenghu site, which include the sky background variations based on moon phase and distance, as well as the airmass calculations to estimate the overall observable time per year for each band.

\subsection{Sky Brightness Model at Lenghu}

WFST is situated on the summit of Mount Saishiteng in Lenghu, at an altitude of 4170\,m, with the geographic coordinates of 38.36$^{\circ}$\,N and 93.53$^{\circ}$\,E. Three years of site monitoring \citep{deng2021lenghu} have revealed that the median seeing is 0.75 arcseconds, the median night temperature variation is 2.4\,degree\,Celsius, and the precipitable water vapor is below 2\,mm for 55\,percent of all night time observations. The atmospheric pressure at the Lenghu site is calculated using the standard atmospheric pressure model \citep{cavcar2000international,tremblin2012worldwide,nolan2010detailed}, giving 
\begin{equation}\label{eq2.7}
  P_{{\rm theo}}=P_0 ( 1 - 0.0065 \frac{h}{T_0})^{5.256}, 
\end{equation}
where $T_0=288.15\,{\rm K}$ and $P_0=1013.25\,{\rm hPa}$  represent the standard temperature and atmospheric pressure at the sea level, respectively. The weather data used in the simulation are from the actual observations at the Lenghu site in the past three years (see Extended Data Fig.~2 of \citet{deng2021lenghu} for details). Focusing on weather conditions only, the total `observable' night times are approximately 2200\,hr in one year, and  roughly 70\% of them are `clear' for observations.

Ground-based astronomical observations are inevitably influenced by background radiation from the atmosphere, which introduces significant photon noise to target observations. The sky radiation originates from the scattering of light from celestial sources and the radiation emitted by the atmosphere itself. The night-sky background mainly consists of scattered moonlight, scattered starlight, zodiacal light, atmospheric thermal radiation and absorption, as well as non-thermal airglow emission. Among these components, scattered moonlight is the brightest natural light source in the night sky and the primary contributor to the optical background noise. Therefore, having a moonlight model to accurately calculate the sky background brightness is crucial.

A moonlight model was developed by \citet{krisciunas1991model}, which determines the brightness distribution of the scattered moonlight. This model takes into account the local extinction coefficient, the zenith distance of the moon, the zenith distance of the target field, and the angular separation between the moon and the target field. However, it should be noted that this model is only empirically calibrated for the $V$ band. \citet{noll2012atmospheric} presented an improved model that allows for predicting the entire optical spectrum of scattered moonlight brightness. The sky background brightness contributed by moonlight can be calculated using two equations based on the Rayleigh scattering and Mie scattering in the atmosphere,  described by \citet{noll2012atmospheric} as
\begin{equation}\label{eq1}
  B_{\rm moon}(\lambda) = B_{\rm moon,R}(\lambda) + B_{\rm moon,M}(\lambda), 
\end{equation}
and
\begin{equation}\label{eq2}
\begin{split}
  B_{\rm moon,R/M}(\lambda) = f_{\rm R/M}(\rho)I^{\ast}t^{X_{\rm moon}}(\lambda)\\ \times(1-t^{X_0}_{\rm R/M}(\lambda)).
\end{split}
\end{equation}
The calculation of Rayleigh scattering and Mie scattering is as follows:
\begin{equation}\label{eq3}
  f_{{\rm R}}(\rho)= 10^{5.70} (1.06 + \cos^2(\rho)), 
\end{equation}
and 
\begin{equation}\label{eq4}
  f_{{\rm M}}(\rho)=10^{7.15 - (\rho/40)}. 
\end{equation}
The moon illuminance is proportional to
\begin{equation}\label{eq5}
\begin{split}
  I^{\ast} \propto 10^{-0.4 (0.026 | \phi|+4.0 \times 10^{-9} \phi^4)}\\ \times (d_{{\rm moon}})^{-2}, 
\end{split}
\end{equation}
where $\rho$ represents the angular separation of the target field from the moon, $d_{\rm moon}$ represents the relative distance to the moon (mean = 1), and $\phi$ can be associated with the moon phase angle $\alpha$ (an angle of separation between the sun and the moon observed from the earth) by $\alpha=180^{\circ}-\phi$. $X$ represents the airmass to be derived using the formula given by \citet{krisciunas1991model} as:
\begin{equation}\label{eq6}
  X=(1 - 0.96 \sin^2 (z))^{-0.5}. 
\end{equation}
Here $z$ is a zenith distance of the target. The transmission $t(\lambda)$ can be estimated by
\begin{equation}\label{eq_add1}
  t(\lambda)= e^{-\tau_0(\lambda)X}=10^{-0.4k(\lambda)X}.
\end{equation}
Due to the dominance of Rayleigh scattering at blue wavelength, there are:
\begin{equation}\label{eq_add2}
\begin{split}
  \tau_R(\lambda) = \frac{P}{1013.25}(0.00864+6.5 \times 10^{-6}H)\\ \times \lambda^{-(3.916+0.074\lambda+0.050/ \lambda)},
\end{split}
\end{equation}
where $P$ represents atmospheric pressure and $H$ represents height. Aerosol scattering becomes as important at red wavelength as Rayleigh scattering, and the aerosol extinction coefficient is parametrised by
\begin{equation}\label{eq_add3}
  k_{{\rm aer}}(\lambda) \approx k_0\lambda^{\alpha},
\end{equation}
where $k_0 = 0.013 \pm 0.002$ mag\,arcsec$^{-2}$ and $\alpha = -1.38 \pm 0.06$, with the wavelength $\lambda$ in ${\rm \mu m}$.

\begin{figure}[htb]
  \centering
  \includegraphics[width=1\columnwidth]{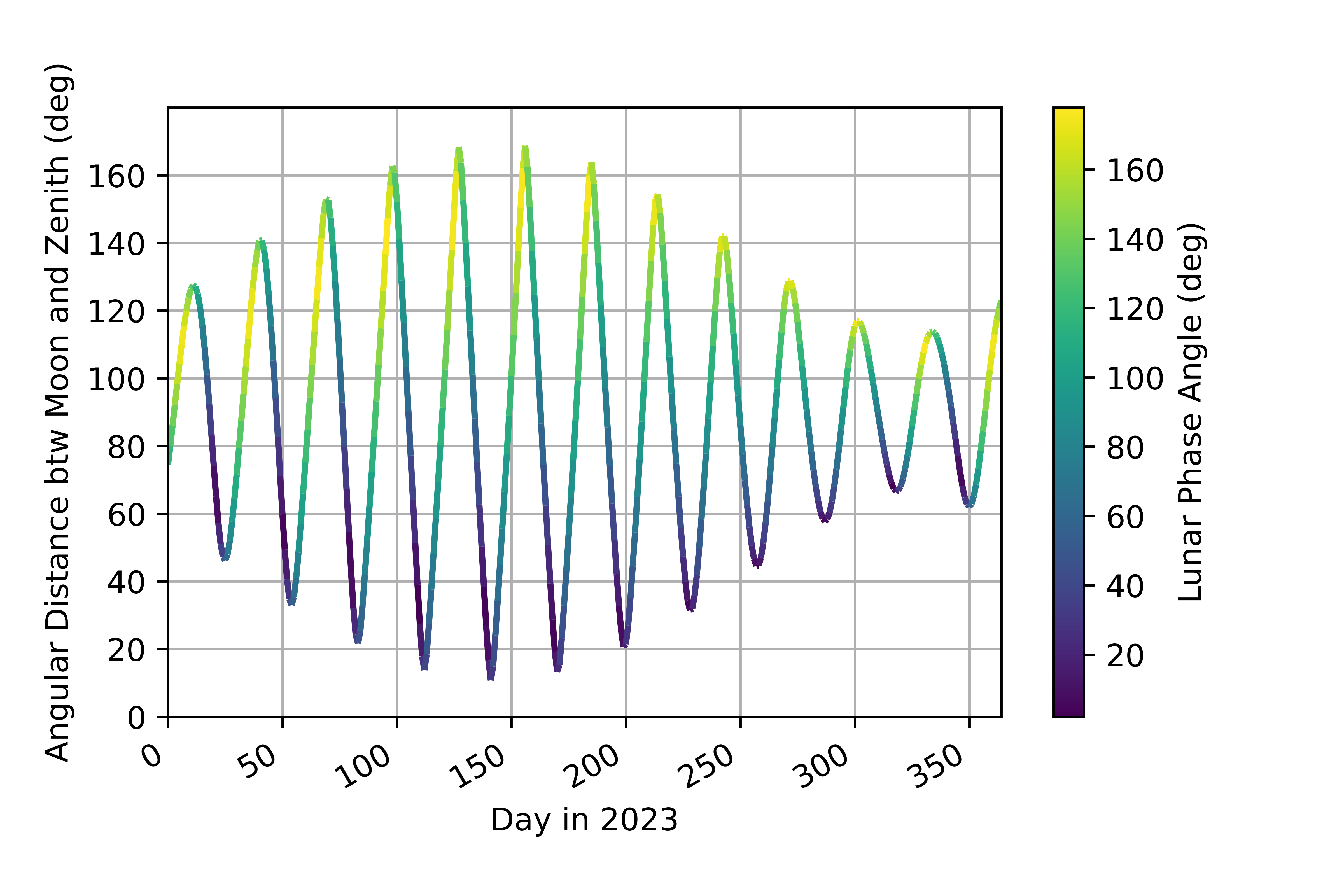}
  \caption{\small Moon phase and zenith distance at the Lenghu site in 2023. The horizontal axis indicates each night of the year, while the vertical axis shows the angular distance between the moon and the zenith at midnight every night. Different colors represent different phases, where 0\,degrees represents the new moon and 180\,degrees represents the full moon.}
  \label{fig1}
  \end{figure}

The variation of the moon phase and position in 2023 is illustrated in Figure~\ref{fig1}. In addition to the moon's influence, the airmass plays a significant role in background radiation and absorption for ground-based observations. Airmass can be calculated through the zenith distance of the pointing. A lower airmass indicates less atmospheric extinction, resulting in higher imaging quality with the same exposure time, particularly for shorter wavelengths.

\subsection{Instrumental Parameters}

During the typical operation of a telescope, the slew time overlaps with the readout time of an exposure. For WFST CCDs, the readout time is 10 seconds (as indicated in Table~\ref{tab1}). Therefore, our analysis primarily focuses on the slew time and the time required for filter changes (5 minutes, also shown in Table~\ref{tab1}). The slew time refers to the duration it takes for the telescope to move from its current pointing position to the next one. Table~\ref{tab1} provides additional instrumental parameters related to the time budget, such as maximum/minimum altitude, tracking accuracy, peak velocity, and peak acceleration and deceleration. Assuming that the telescope experiences constant acceleration and deceleration and has a maximum speed limit, we can calculate the slew time based on the distance between two positions. If the distance allows for the maximum speed, we can compute the peak velocity and telescope positioning time using the following formula \citep{delgado2014lsst}:
\begin{equation}\label{eq2.1}
  V_{\rm peak}=\sqrt{\frac{2 |Distance|}{\frac{1}{accel}+\frac{1}{decel}}}
\end{equation}
and
\begin{equation}\label{eq2.2}
  delay = \frac{V_{\rm peak}}{accel} + \frac{V_{\rm peak}}{decel}. 
\end{equation}
If the distance between two pointings is large enough, a telescope will reach its maximum moving speed $V_{\rm max}$ (determined by the performance of the telescope), the slew time of a telescope can be derived using the following formula \citep{delgado2014lsst}:
\begin{equation}\label{eq2.3}
  delay = \frac{V_{\rm max}}{accel} + \frac{d2}{V_{\rm max}} + \frac{V_{\rm max}}{decel}, 
\end{equation}
where
\begin{equation}\label{eq2.4}
  d2 = |Distance|-\frac{V_{\rm max}^2}{2 accel}-\frac{V_{\rm max}^2}{2 decel}
\end{equation}
Here, $accel$ and $decel$  represent the acceleration and deceleration of the telescope, respectively. By employing equations (8)--(11), we can calculate the slew time for specific telescope cases. The total time for a single-band observation is then the sum of the exposure time and the slew time. Alternatively, if the slew time is shorter than the readout time, it becomes the sum of the exposure time and the readout time. We note that the real overhead time would be longer than the simple assumption adopted here. For example, communications and responses among different systems (camera, filter system, shutter, and telescope) may lead to non-negligible time during the telescope positioning. Also, time for filter exchange needs to be considered when switching the filter for the subsequent exposure.

%有些参数可能是有问题的，先与相机团队确认。
\begin{table*}[htb]
\begin{center}
\caption[]{Basic parameters of WFST}\label{tab1}
\setlength{\tabcolsep}{10mm}{
    \begin{tabular}{lc}
    \hline\noalign{\smallskip}
    Min elevation angle & 10 deg          \\
    Max elevation angle & 90 deg               \\
    Blind pointing & $\leq5''$ rms               \\
    Tracking accuracy & $\leq0.1''$ rms     \\
    Camera readout & 10 sec                          \\
    Filter change time & 2 min   \\
    Max mounted filters & 6 \\
    Peak velocity & $\geq2~{\rm deg/s}$\\
    Peak acceleration and deceleration & $\geq1~{\rm deg/s}^2$\\
    \hline\noalign{\smallskip}
\end{tabular}}
\end{center}
\end{table*}

\subsection{Simulation Results}

\begin{figure*}[htb]
	\centering
  \begin{minipage}[c]{0.45\textwidth}
  \centering
  \includegraphics[width=\textwidth]{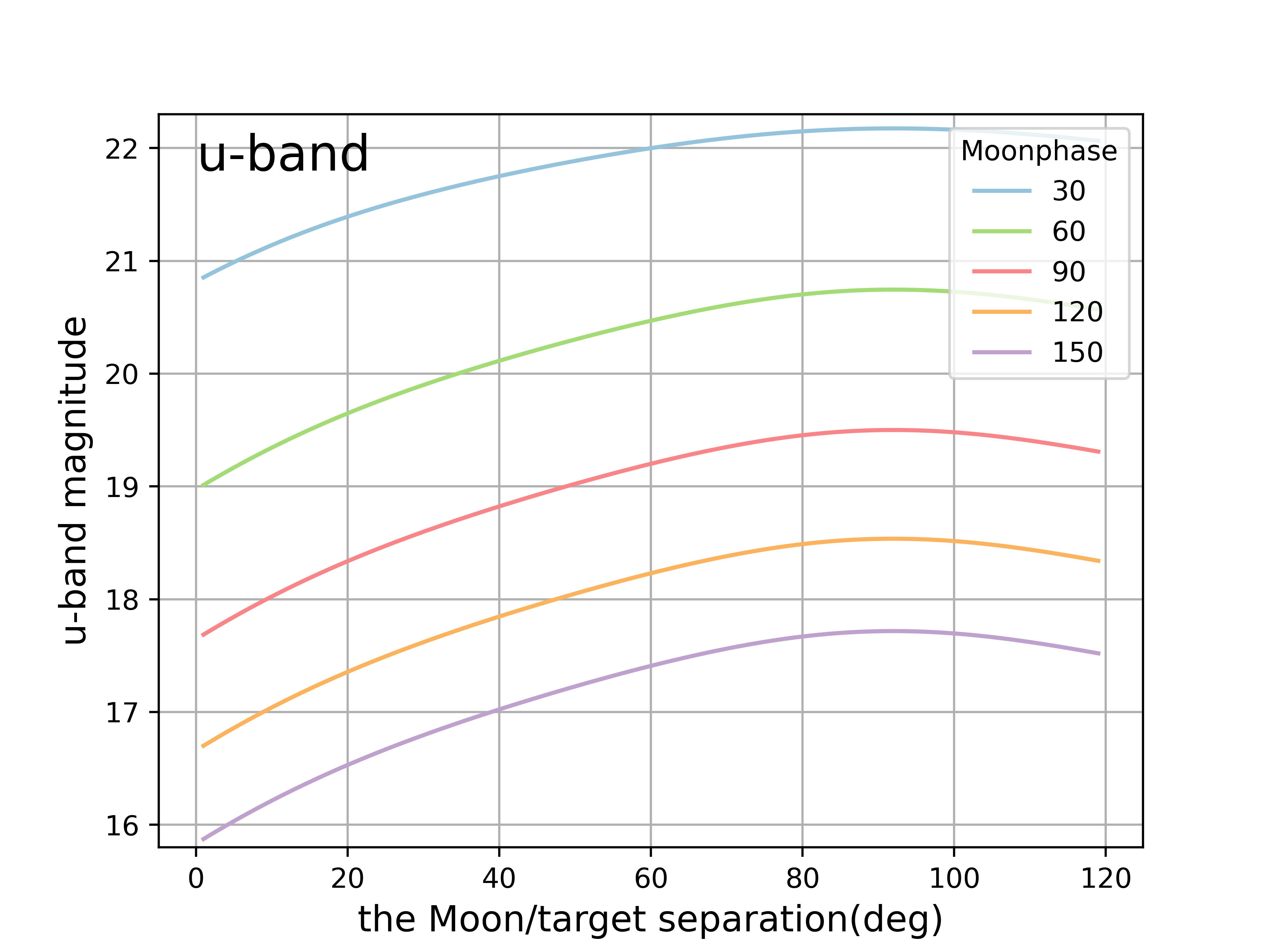}
  \end{minipage}
  \begin{minipage}[c]{0.45\textwidth}
  \centering
  \includegraphics[width=\textwidth]{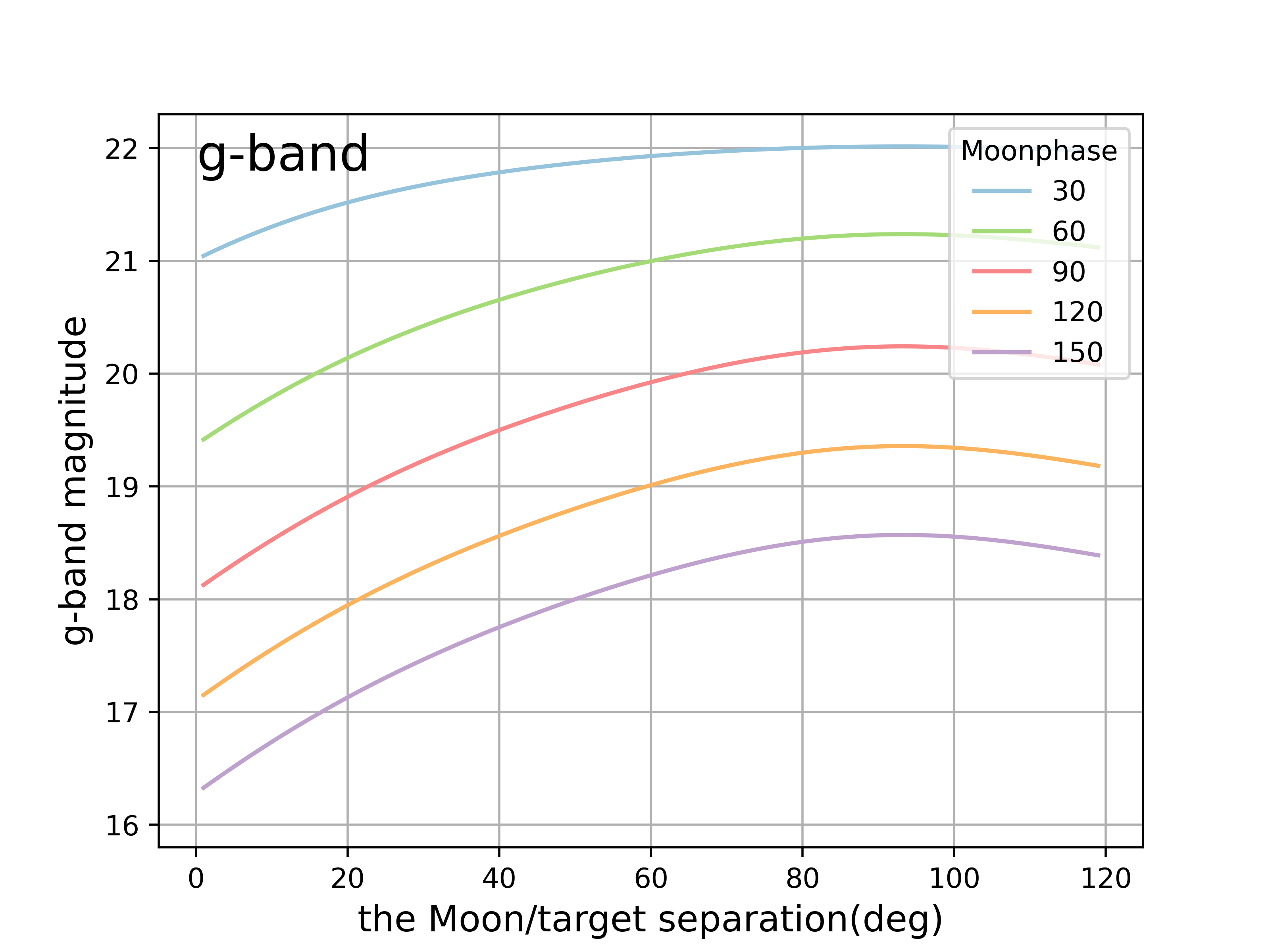}
  \end{minipage}

  \begin{minipage}[c]{0.45\textwidth}
  \centering
  \includegraphics[width=\textwidth]{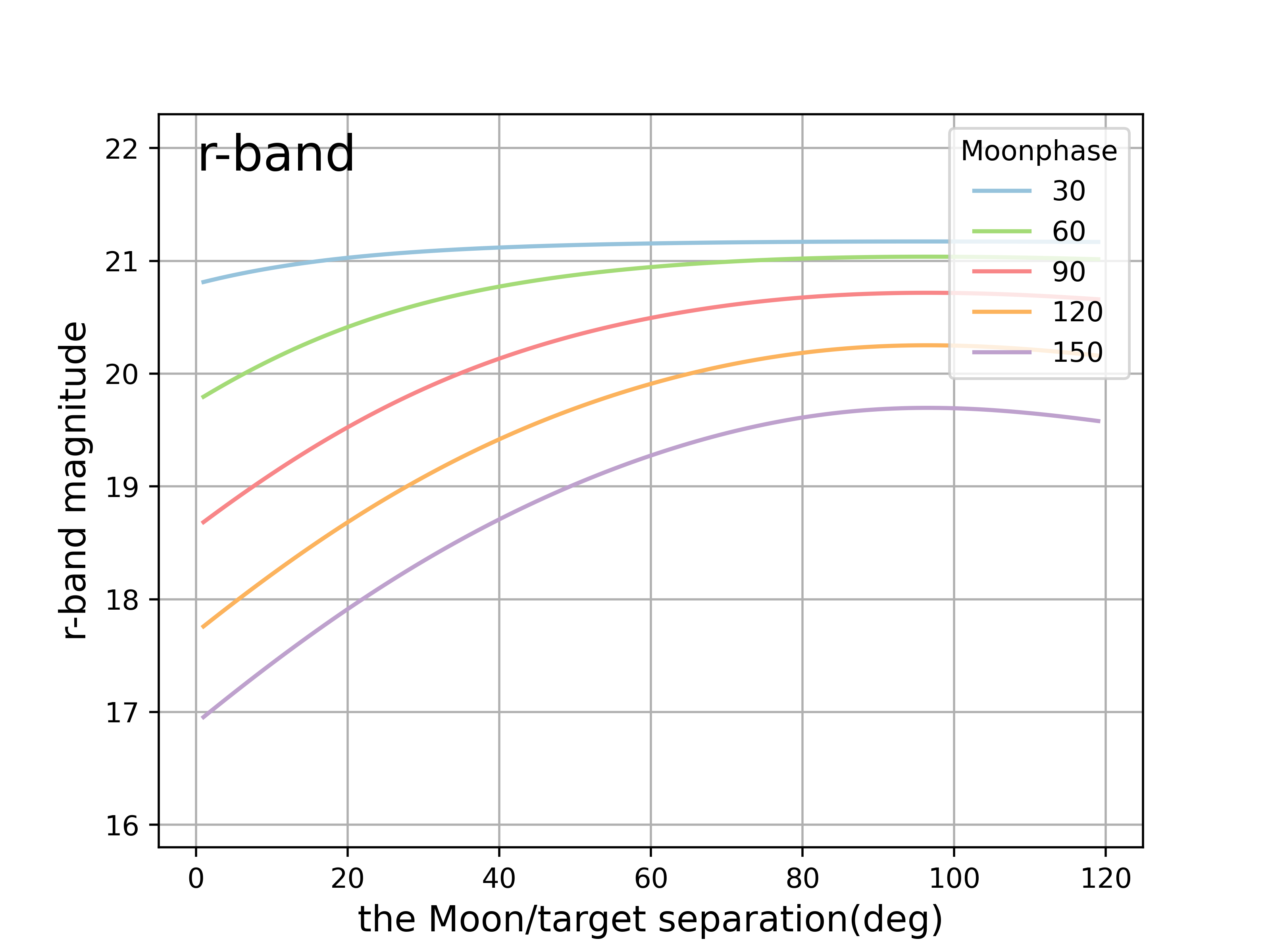}
  \end{minipage}
  \begin{minipage}[c]{0.45\textwidth}
  \centering
  \includegraphics[width=\textwidth]{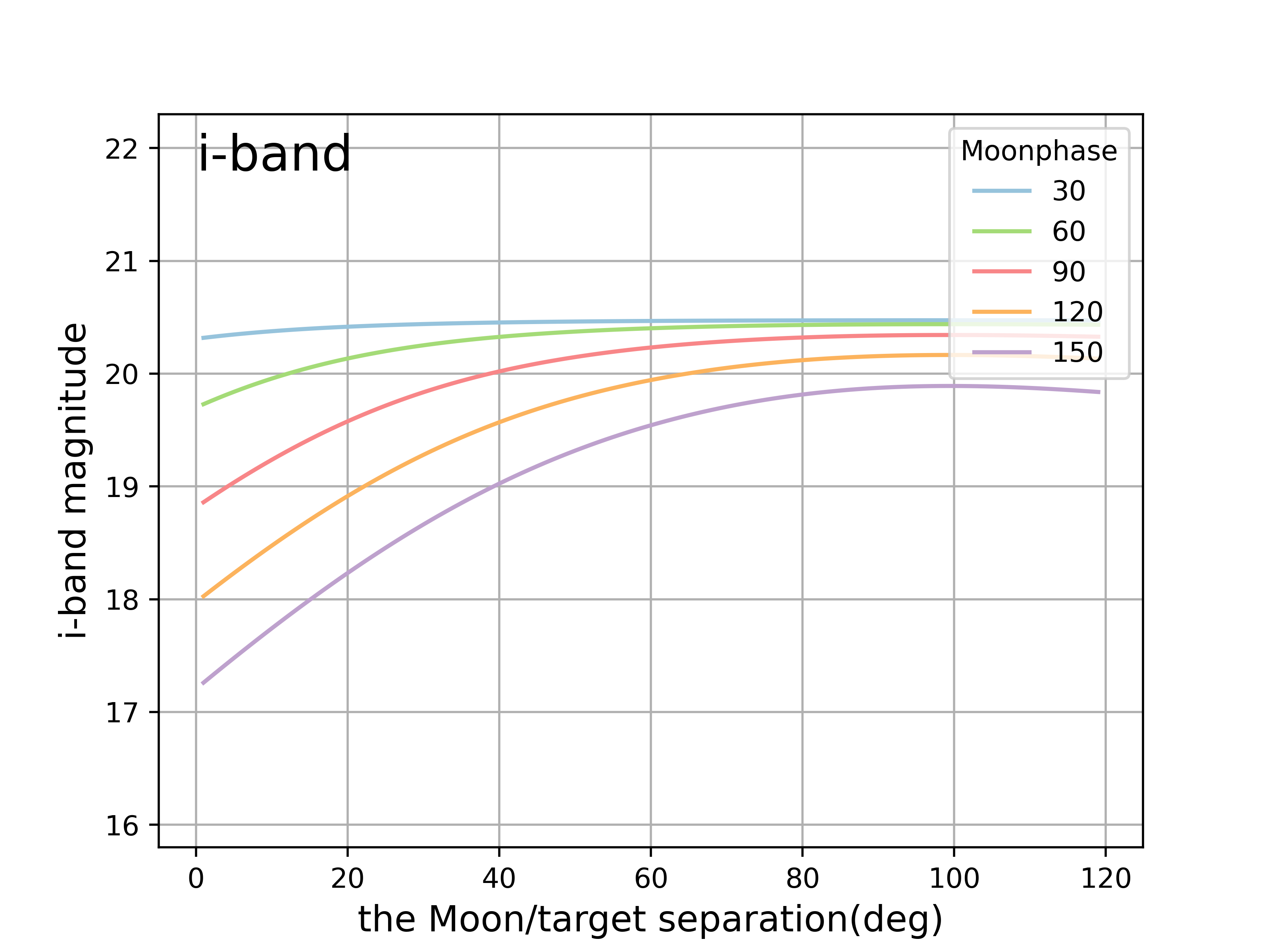}
  \end{minipage}

  \begin{minipage}[c]{0.45\textwidth}
  \centering
  \includegraphics[width=\textwidth]{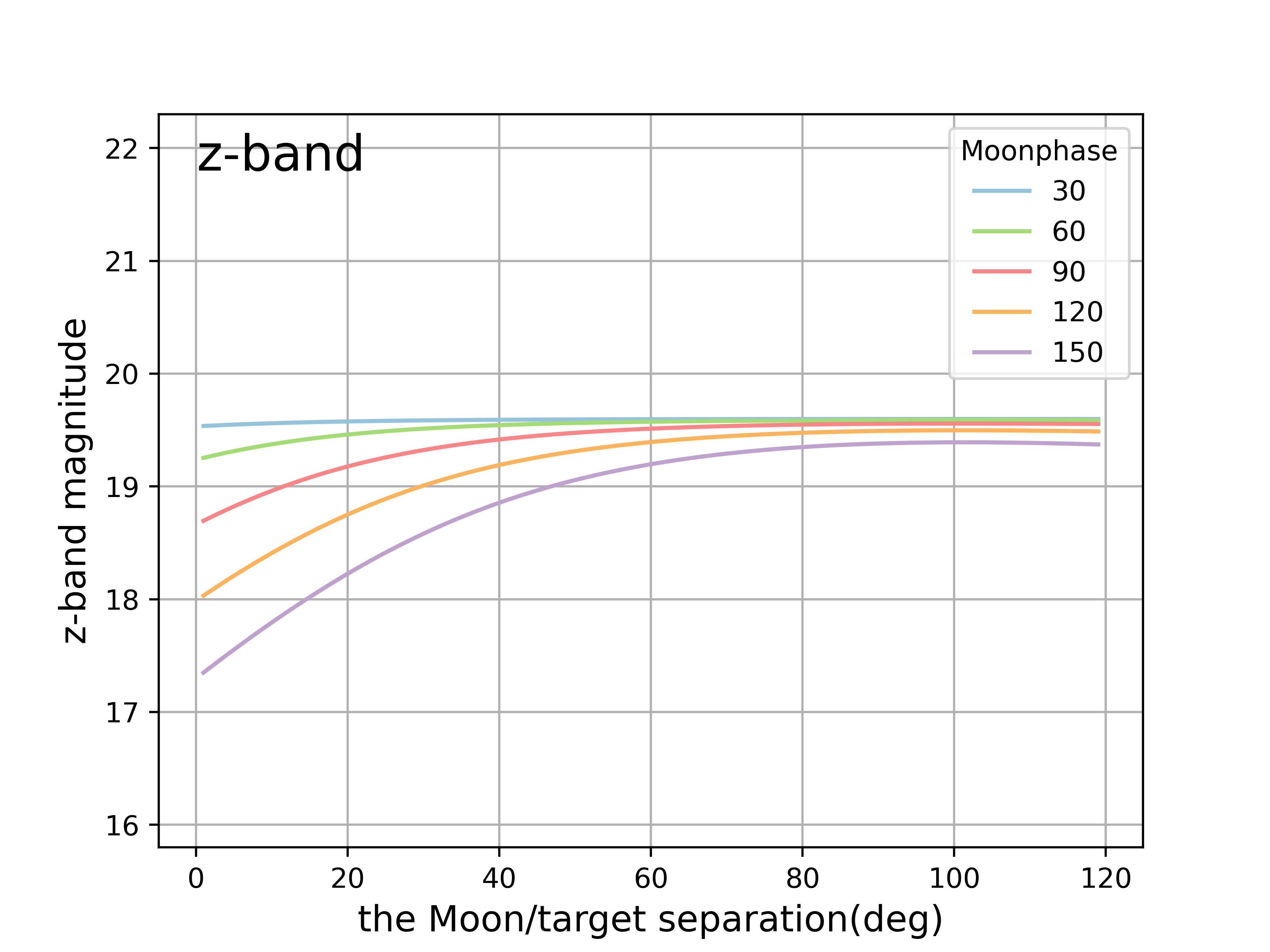}
  \end{minipage}
  \begin{minipage}[c]{0.45\textwidth}
  \centering
  \includegraphics[width=\textwidth]{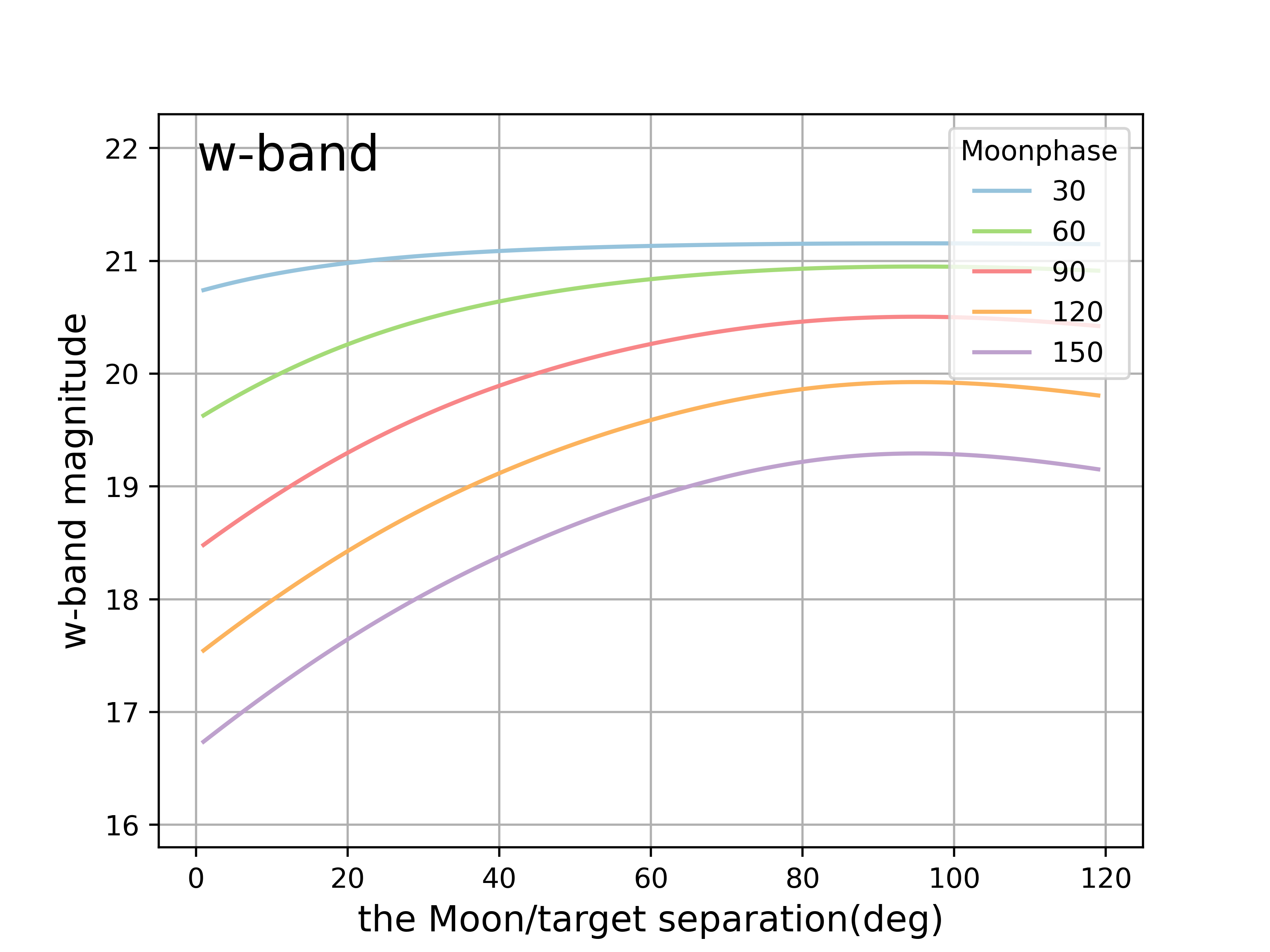}
  \end{minipage}
	\caption{Relationships between sky background and angular separation between moon and target field for the opitcal $u,g,r,i,z,w$-bands. Different colors denote different phases of the moon. The zenith distance of the moon, target region, and the sky background are set to $60{\degr}$, $40{\degr}$, and 22.3\,${\rm mag\,arcsec}^{-2}$, respectively.}
  \label{fig2}
\end{figure*}

The blue optical and near ultraviolet (NUV) bands are significantly affected by the moon phase and airmass. Therefore, it is crucial to reduce the impact of these factors, so as to achieve a specific depth and minimize exposure time. By utilizing a calculation model for the moon's contribution to sky brightness, we can assess how brightness changes in different bands and sky regions under various moon phases. Here, we assume a target field zenith distance parameter of 40\,degrees and a moon zenith distance of 60\,degrees. The results of the sky background as a function of moon phase are depicted in Figure~\ref{fig2} (Considering a more intuitive display of the impact of moon phase, a constant moonless sky background has been added to all bands). From the figure, it is apparent that in optical observations, the impact of the moon becomes more pronounced as the wavelength becomes shorter. With the exception of the $u$ and $g$ bands, the sky background brightness does not exhibit significant variation with the distance between the target field and the moon when the moon phase is around 30\,degrees. Hence, in grey nights, the influence of the moon phase in optical observations can be largely disregarded, except for the $u$ and $g$ bands.

%\subsubsection{Distribution of accessible observing time in the whole sky}
The signal-to-noise ratio (S/N) for a CCD image can be quantified as follows
\begin{equation}
\label{eq10}
  \frac{S}{N} = \frac{N_{\ast}}{\sqrt{N_{\ast} + n_{\rm pix}(N_{\rm S} + N_{\rm D} + N_{\rm R}^2)}},
\end{equation}
where $N_*$ is the total number of photon electrons collected from a celestial object per exposure time, $n_{\rm pix}$ is the number of pixels under consideration for the S/N calculation, $N_{\rm S}$ is the total number of photoelectrons collected from the sky background per exposure time, $N_{\rm D}$ is the total number of electrons from the dark current, and $N_{\rm R}$ is the total number of electrons from the readout noise. If the total noise $\sqrt{N_{\ast} + n_{\rm pix}(N_{\rm S} + N_{\rm D} + N_{\rm R})}$ is dominated by sky background noise, then the formula for calculating the signal-to-noise ratio of CCD images can be written as \citep{howell2006handbook}:
\begin{equation}\label{eq11}
  \frac{S}{N} = \frac{N_{\ast}}{\sqrt{n_{\rm pix} N_{\rm S}}}.
\end{equation}
Using the above equation and combining it with the following magnitude formula:
\begin{equation}\label{eq12}
  m_1 - m_2 = -2.5 log_{10}(\frac{f_1}{f_2}).
\end{equation}

It is evident that the signal-to-noise ratio (SNR) decreases by 40--50\% when the sky background at the target field is 1.5--2 mag brighter. As the $u$ band is most sensitive to the moon phase, we set the threshold for this band to be 2\,mag\,arcsec$^{-2}$ brighter than the background without moonlight. For the other bands, the thresholds are set to be 1.5\,mag\,arcsec$^{-2}$ brighter than the background without moonlight in subsequent simulations. Since there is a lack of multiband sky background information at Lenghu, we utilize the moonless night-sky brightness data from Cerro Pach$\acute{\rm o}$n as $u$ (22.99), $g$ (22.26), $r$ (21.20), $i$ (20.48), $z$ (19.60) in units of AB\,mag\,arcsec$^{-2}$ \citep{ivezic2019lsst}. Consequently, the thresholds for each band are as follows:  $u$ (20.99), $g$ (20.76), $r$ (19.70), $i$ (18.70), and $z$ (18.10).

By integrating four key factors - site, meteorology, slew time, and sky background - we can simulate whether a specific target area is suitable for observation at a given time. Additionally, we can sample and calculate the suitability of all sky areas for observations every five minutes during observable nights. After performing these calculations for an entire year, we can generate a map showing the maximum observable time in each band for various sky areas. The following observation conditions are considered: (1) Airmass must be less than 1.5, corresponding to a zenith distance of each sky area being less than 49.52\,degrees. (2) A sky area is deemed `observable' only if its sky background is darker than the specified threshold. (3) Weather conditions based on \citealp{deng2021lenghu} are taken into account. We use the obtained real weather conditions table for the whole year of 2021, which records whether each night and time in 2021 is suitable for observation at intervals of one minute. Therefore, the calculation of the maximum observable time only selects the time when the weather conditions are suitable for accumulation.

\begin{figure*}[htb]
	\centering
  \begin{minipage}[c]{0.45\textwidth}
  \centering
  \includegraphics[width=\textwidth]{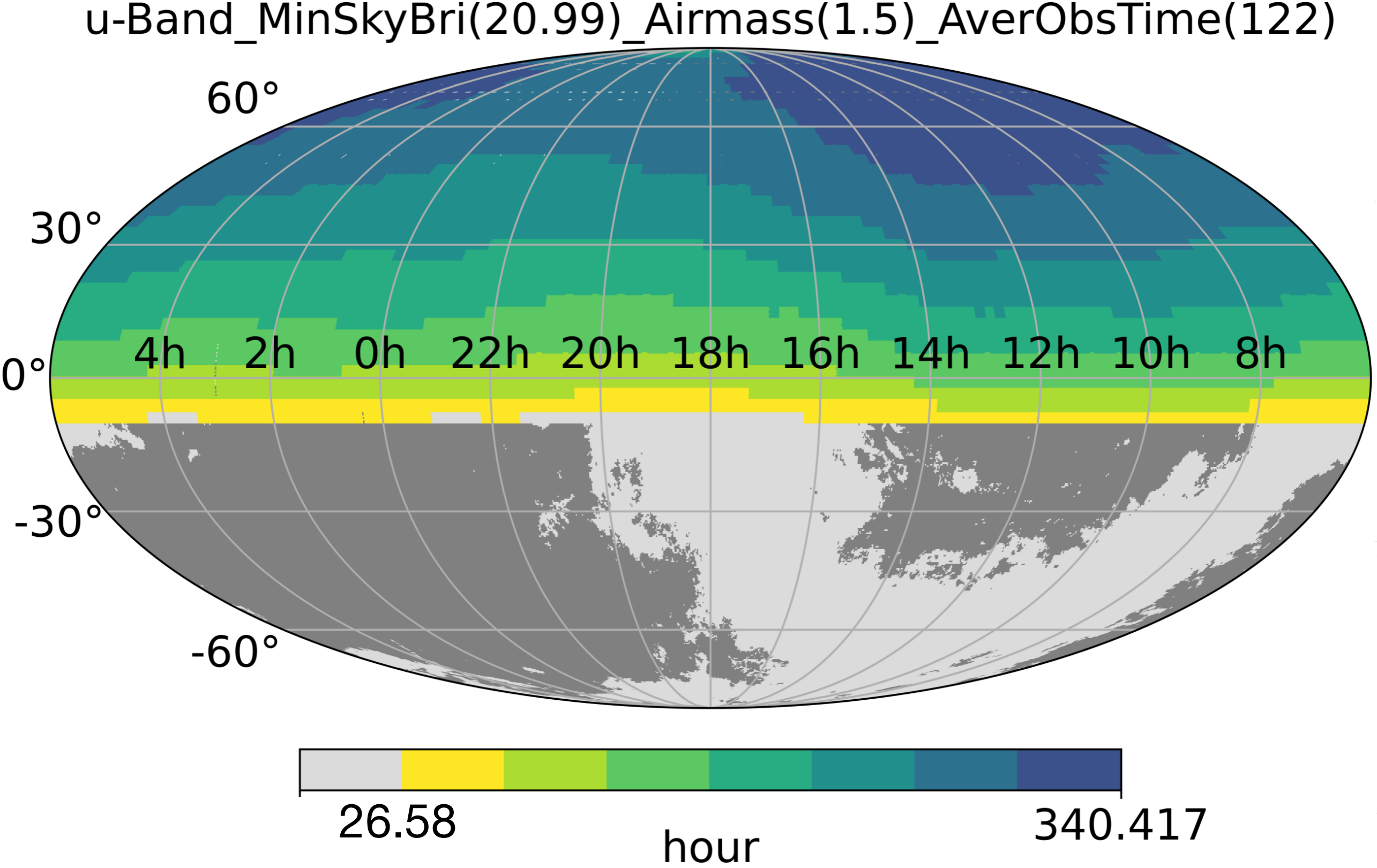}
  \end{minipage}
  \begin{minipage}[c]{0.45\textwidth}
  \centering
  \includegraphics[width=\textwidth]{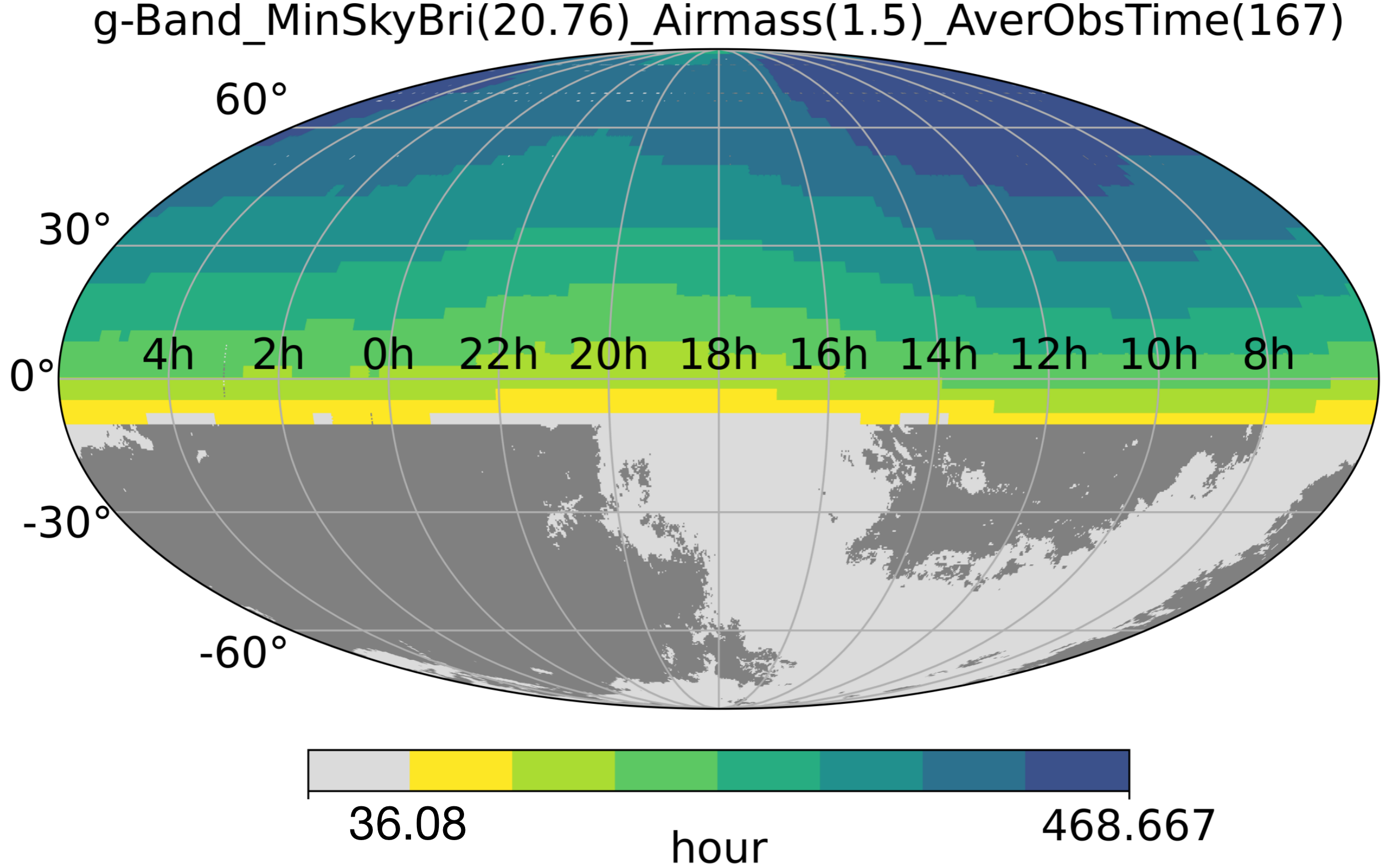}
  \end{minipage}

  \begin{minipage}[c]{0.45\textwidth}
  \centering
  \includegraphics[width=\textwidth]{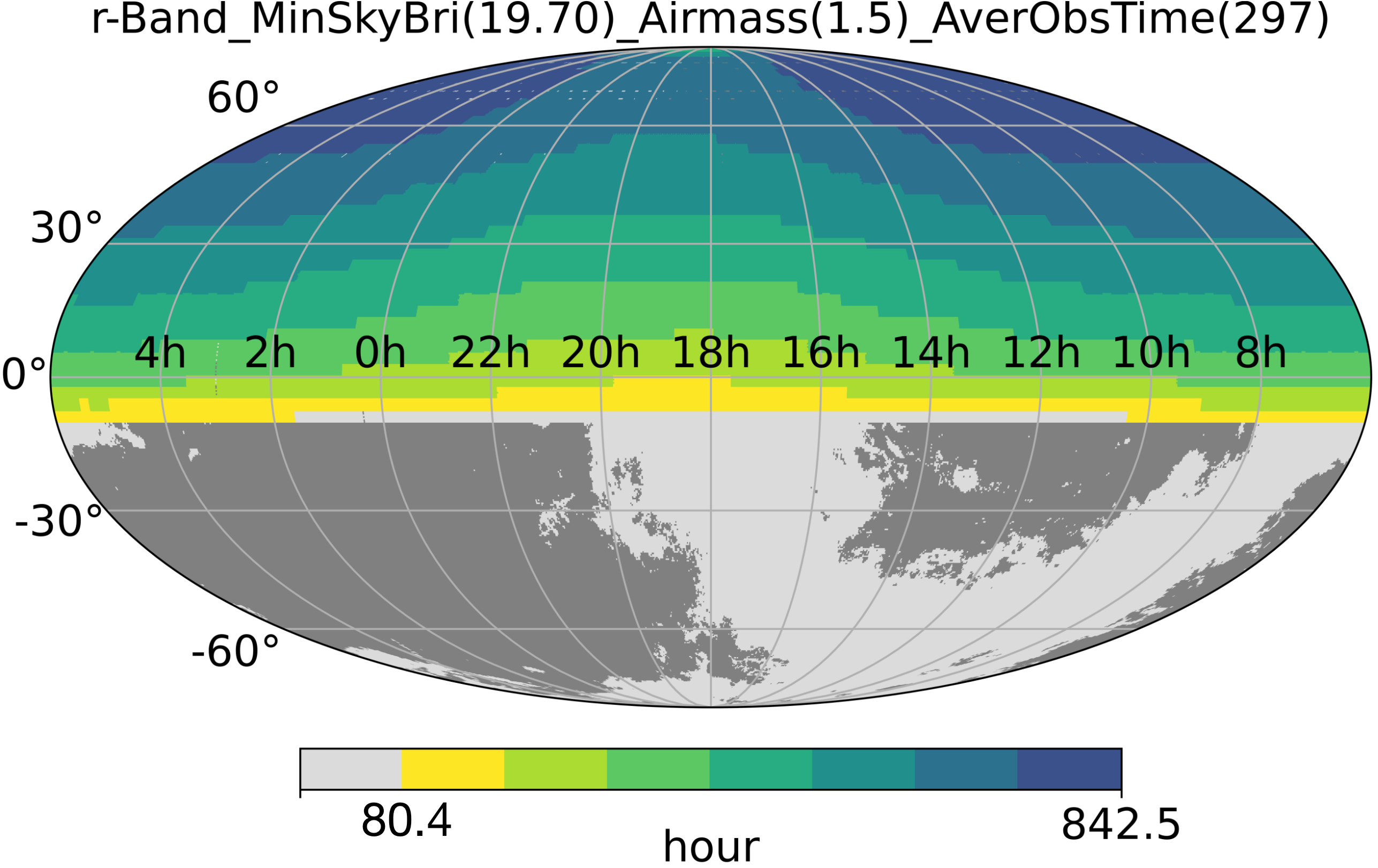}
  \end{minipage}
  \begin{minipage}[c]{0.45\textwidth}
  \centering
  \includegraphics[width=\textwidth]{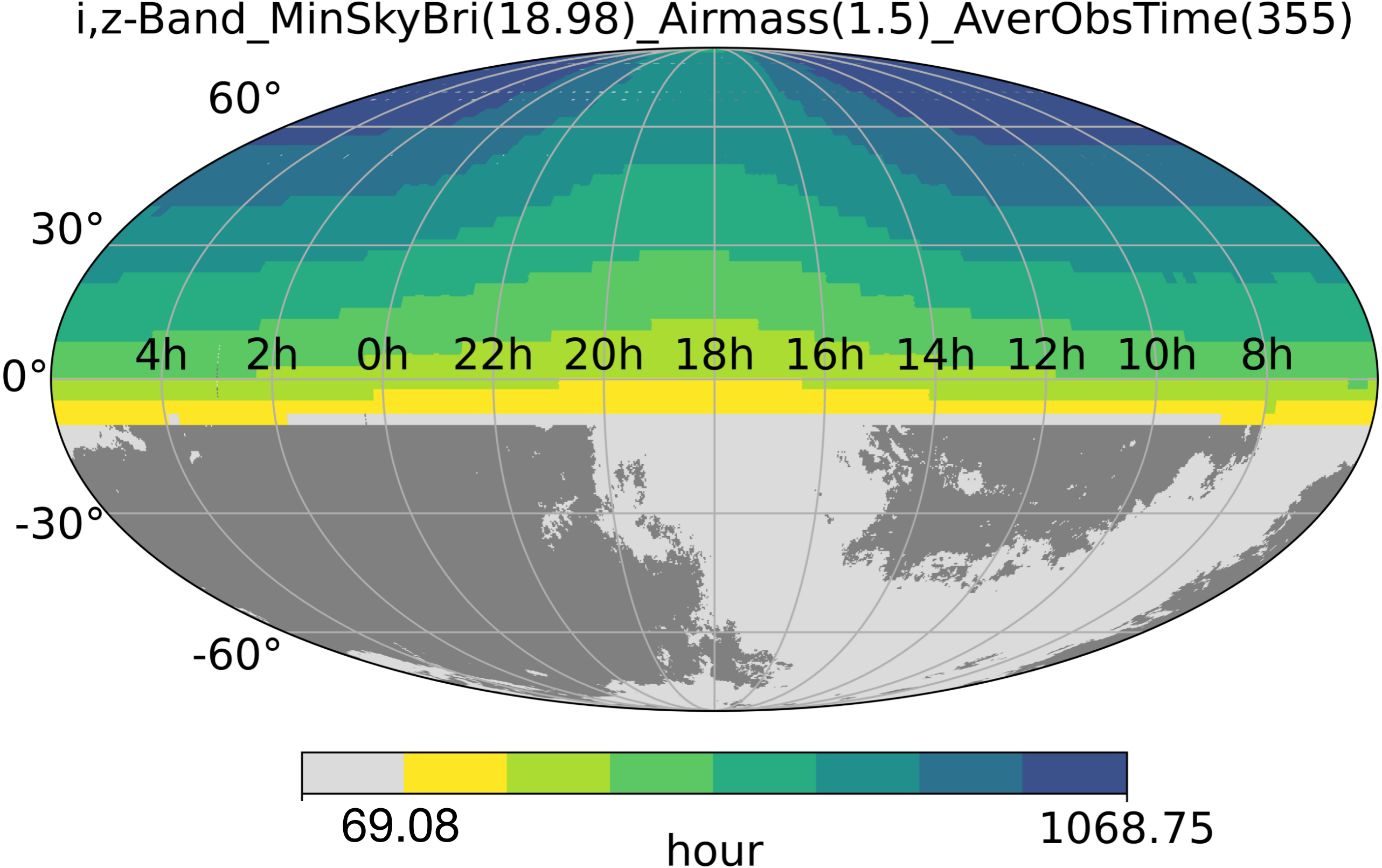}
  \end{minipage}
\caption{observable time distribution of the entire accessible sky, the title of each plot shows the band, threshold value of sky background, airmass, and average observable time over the entire sky area, respectively.}
  \label{fig3}
\end{figure*}

The maximum observable time for each sky area in each band is summarized in Figure~\ref{fig3} (Since the simulation results of the $i$ band and $z$ band are similar, the results of these two bands are placed on the same graph). We observe that high declination sky areas often have a larger observable time. This can be attributed to these areas being more likely to meet the airmass constraint, which is influenced by the geographical location of the Lenghu site. Moreover, the uneven distribution of accessible observable time across different Right Ascensions is influenced by seasonal and weather-related factors. Winter nights generally offer much longer observable time compared to summer nights. Notably, $u$-band observations exhibit the shortest observable time, while $i$ and $z$ bands have the longest average observable time, given that the $u$ band is more sensitive to moon conditions.

\section{Survey Configuration and Scheduling}
\label{sect:3}

% \subsection{Tiling}

The basic observation unit is defined as a single pointing with specific exposure time, and the field of view of the pointing determines the size of a unit grid, we divide the sky into basic observation units called tiles, each covering approximately 6\,deg$^2$, which is comparable to the effective Field of View of WFST. Consequently, the sky is divided into numerous tiles that cover regions of interest. The center of each tile can coincide with the center of the telescope pointing. To generate a coordinate list that covers a series of tiles in a specific order is a major task of the scheduling design. 

\subsection{Sky Division for WFST: Basic Mode}

\begin{figure*}[htb]
  \begin{subfigure}[c]{0.495\linewidth}
  \centering
  \includegraphics[width=60mm]{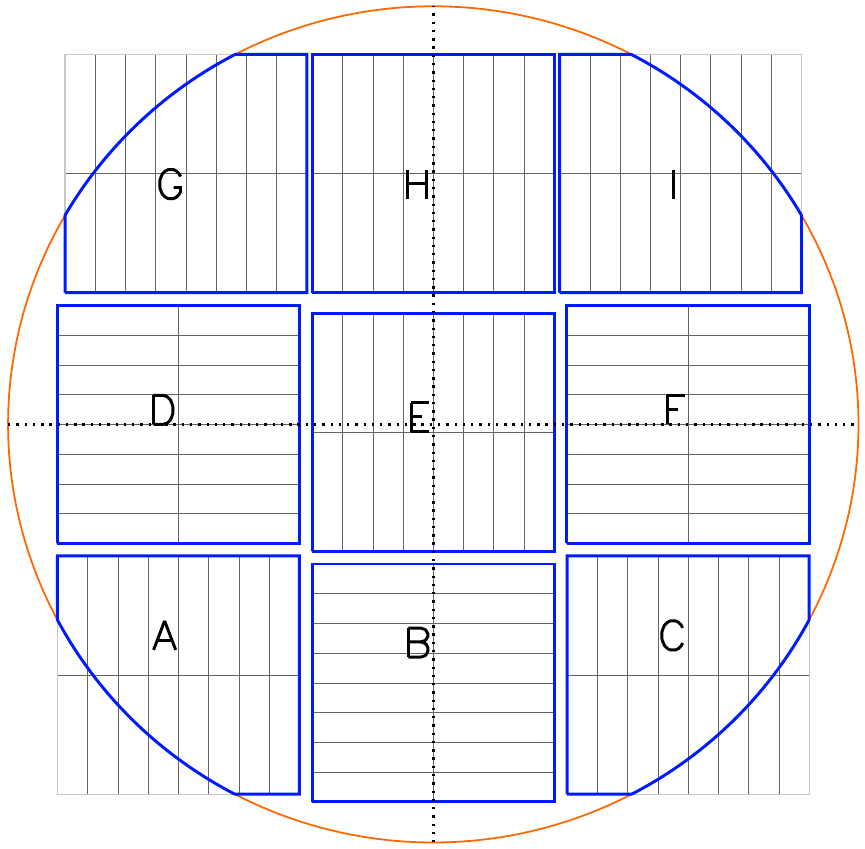}
  \caption{}
  \label{fig5.1}
  \end{subfigure}%
  \begin{subfigure}[c]{0.495\textwidth}
  \centering
  \includegraphics[width=60mm]{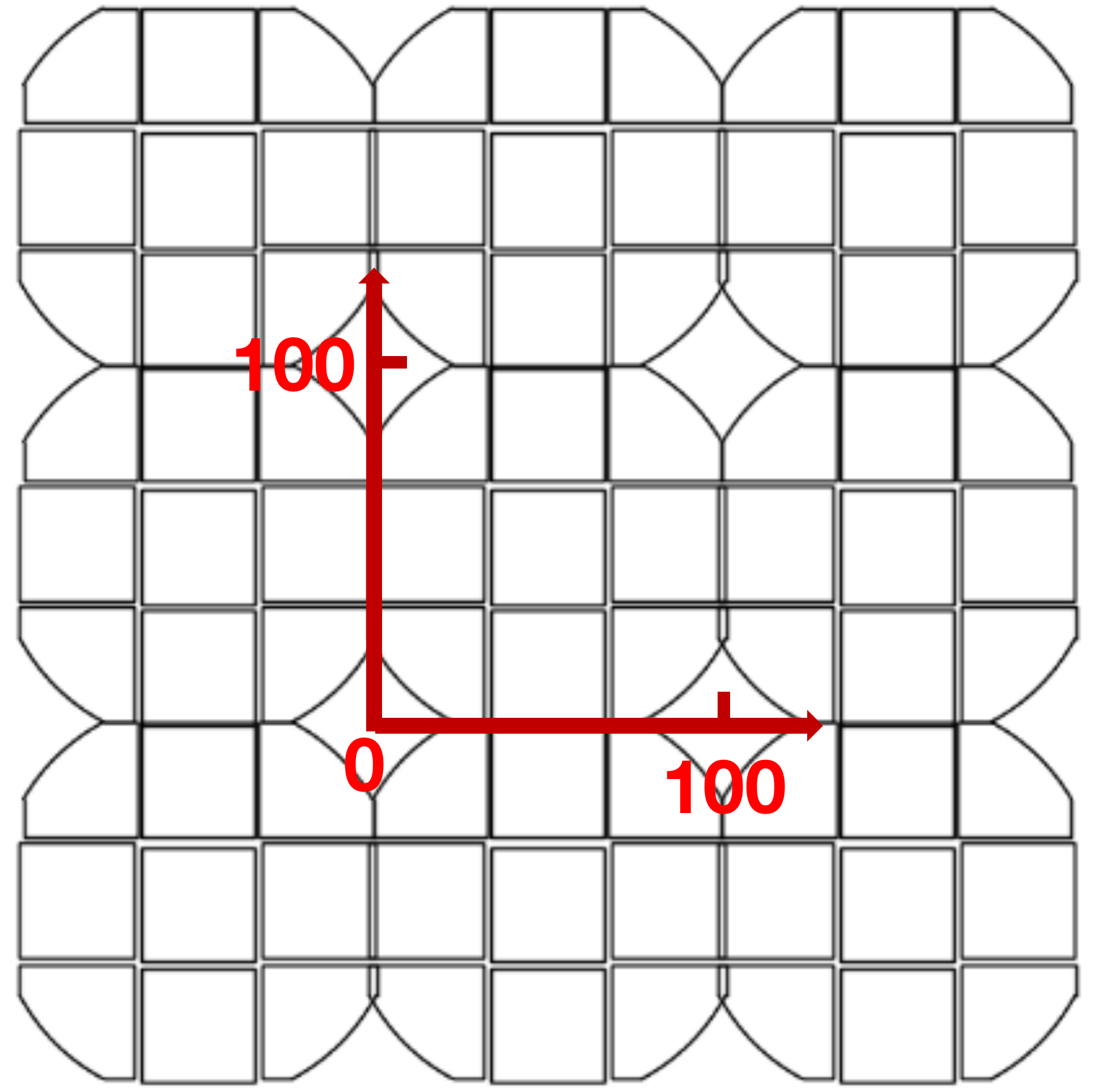}
  \caption{}
  \label{fig5.2}
  \end{subfigure}%
  \caption{Left (a): the layout of the focal plane of WFST, nine squares represent nine $9{\rm K} \times 9{\rm K}$ CCDs forming a scientific imaging array, and there are four CCDs whose corners are outside the 3-degree field of view. Right (b): layout display of nine focal plane arrays in low declination regions. A red coordinate axis is used to optimize the dithering pattern, which will be introduced in the following text.}
  \label{fig5}
\end{figure*}

The arrangement diagram of the focal plane array of WFST is illustrated in Figure~\ref{fig5.1}. The nine CCDs are denoted as `A' to `I', with each having 16 channels. A single CCD contains $9216 \times 9232$ pixels with a pixel scale is 0.33". The layout direction is depicted in Figure~\ref{fig5.1}. However, a section of four CCDs in the corners falls outside the 3-degree field of view, rendering them incapable of producing useful scientific image information. Furthermore, there are gaps of varying sizes between the CCDs. The gaps between CCDs are mostly 4.86\,mm, while the largest gap is 7.90\,mm between `E' and `H'. As a result, WFST cannot treat the smallest unit of sky division as a regular hexagon. To address this, we divide the entire sky into small rectangles using a rectangular concatenation method. The schematic diagram of the splicing of nine focal plane arrays is shown in Figure~\ref{fig5.2}. Note that the three sides of the focal plane array are not perfectly aligned. Specifically, CCDs D, B, and F are arranged slightly outward compared to the other five CCDs (as shown in Figure~\ref{fig5.1}), resulting in overlaps in both the X and Y directions when the focal plane arrays are adjacent to each other (as depicted in Figure~\ref{fig5.2}). In addition, at the high declination, the overlap becomes larger at the top side and smaller at the bottom side. We ensure a minimal overlap of 50 pixels so that the the sky area can be continously covered by our survey observations. The final minimum unit of sky division is defined as a rectangle with a length of 2.654\,degrees and a width of 2.597\,degrees, corresponding to a total area of 6.892\,deg$^2$. 

\begin{figure*}[htb]
  \centering
  \includegraphics[width=1\columnwidth]{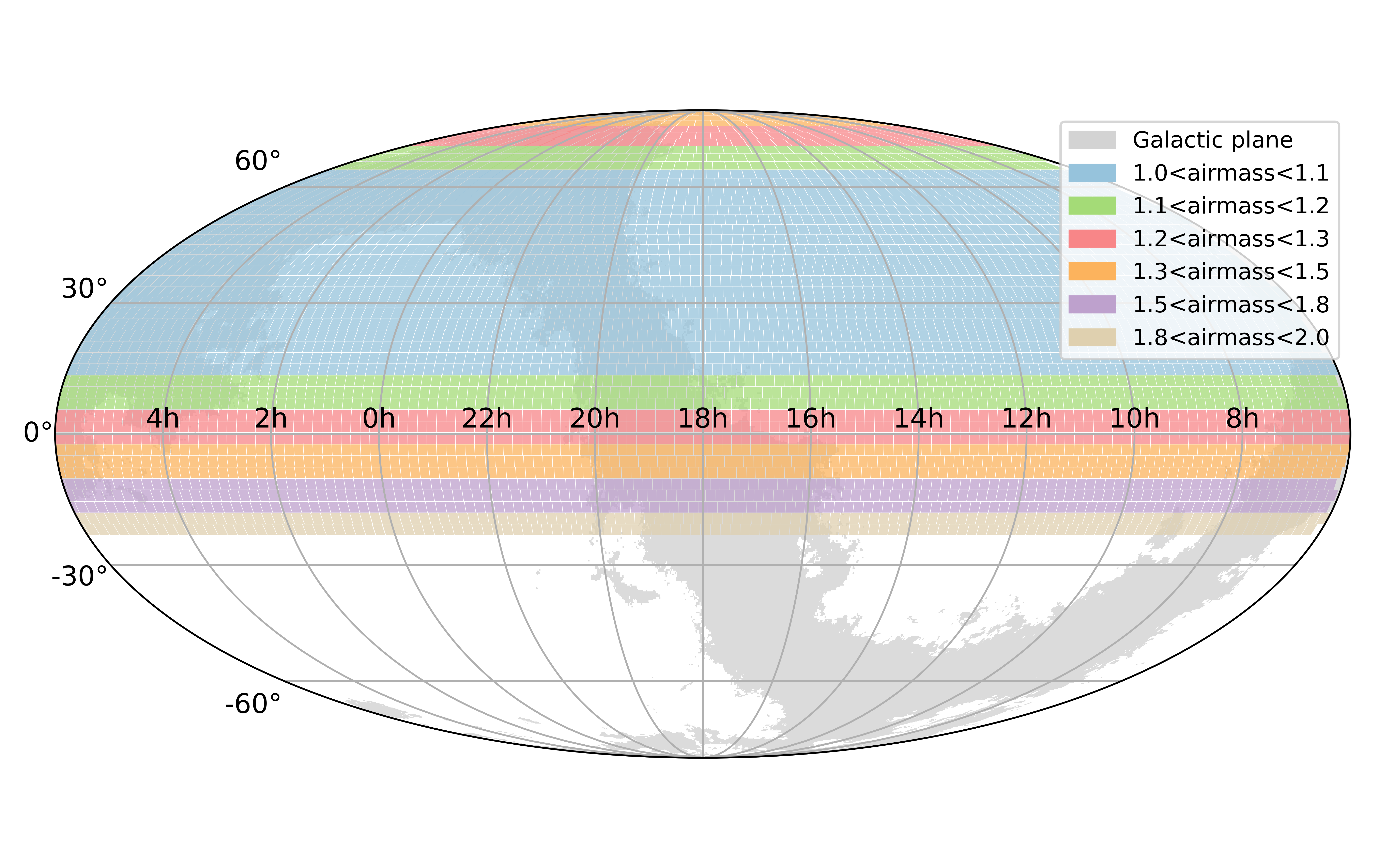}
  \includegraphics[width=1\columnwidth]{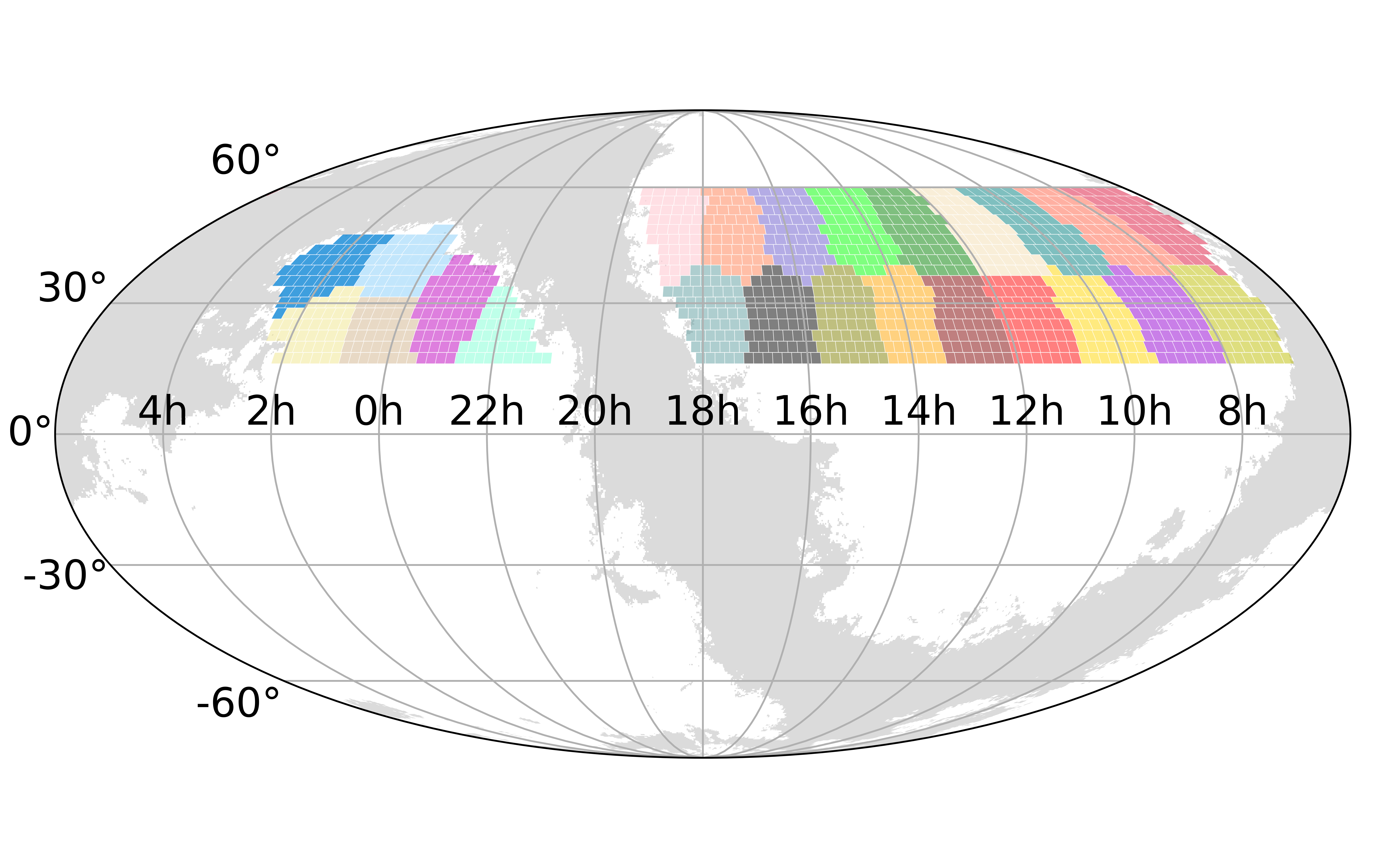}
  \caption{\textbf{Left}:A sky division mode of WFST. the gray area indicates color excess $E_{B-V}$ is greater than 0.15. The best airmass level that each tile can reach is shown by different colors, which is determined by the geographical location of the Lenghu site. There are 3458 tiles with airmass less than 1.5 in total, corresponding to 24,159\,deg$^2$. \textbf{Right}: 24 tile groups are shown in different colors.}
  \label{fig6}
  \end{figure*}

Figure~\ref{fig6} illustrates the basic mode of sky division. Each minimum sky division unit represents a tile, as defined at the beginning of this section, and is assigned an ID number. The tiles are numbered in order from the North to the South Celestial Pole, along the positive direction of the right ascension. The position of the tile center (right ascension and declination) is also recorded along with the ID. 

% The log information of each exposure is stored in specific tables of a database. Table~\ref{tab2} provides detailed descriptions of the relevant tables.

% \begin{table*}[htb]
%   \bc
%   \begin{minipage}[]{100mm}
%   \caption[]{Table of database\label{tab2}}\end{minipage}
%   \setlength{\tabcolsep}{10mm}
%   \small
%     \begin{tabular}{lc}
%     \hline\noalign{\smallskip}
%   Table &  Description \\
%     \hline\noalign{\smallskip}
%   Fields& Stores the central coordinates and IDs of all fields\\
%   Weather& A seeing table and a cloud table at the site\\
%   Configuration tables& Store the configuration parameters for each observing run\\
%   Observation history& All simulated observations\\
%   Slew history& Record the telescope slew time of each visit\\
%     \noalign{\smallskip}\hline
%   \end{tabular}
%   \ec
%   \end{table*}

For two major survey programs of WFST: a wide-field imaging survey and a deep high-cadence survey, sky with significant Milky Way extinction is excluded. Low galactic latitude areas with a color excess $E_{B-V}$ greater than 0.15 are colored in grey in Figure~\ref{fig6}. When applying a upper limits of airmass of 1.5 and Milky Way color excess $E_{B-V}$ of 0.15, repectively, resulting in 2360 remaining tiles with a total area of 15,916 ${\rm deg}^2$. After removing low galactic latitude regions with color excess $E_{B-V}$ greater than 0.1 and 0.2, the remaining tiles are 2097 and 2524, corresponding to a total area of 14144 and 17,013 ${\rm deg}^2$, respectively. The selection of different galactic latitude regions will result in different survey areas, which can be adjusted based on scientific motivations. In this article, we focus on high galactic latitude regions with $E_{B-V}$ smaller than 0.15.

\subsection{Sky Division for WFST: Group Mode}

Assuming a survey covering a total sky area of approximately 8,000\,deg$^2$ as suggested by the WFST wide survey \citep{wfst2023}, we select sky region with declination between 15 and 60\,degrees and $E_{B-V}$ smaller than 0.15 for simulation. There are a total of 1186 tiles. If all 1186 tiles are used for survey scheduling, the computational workload will be particularly high. Therefore, we have used the classic unsupervised learning clustering algorithm known as the K-Means algorithm \citep{macqueen1967classification}. This algorithm divides the tiles into 24 groups\footnote{It takes about half an hour for a single group of survey simulation, which is more convenient to produce simulation results. Other group combinations are also avilable.}, each group is called a `tile group'. As shown in the right panel of Figure~\ref{fig6}, each color block indicates a tile group including approximately 50 tiles. For the following scheduling simulations, we only consider the arrangement of the observation sequence of the 24 tile groups.

%\begin{figure}[htb]
%  \centering
%  \includegraphics[width=1\columnwidth]{two_layer_grid.png}
%  \caption{Secondary sky Division of WFST. The gray area meets the requirement that the color excess $E_{B-V}$ is greater than 0.15, the declination is between 15\,degrees and 60\,degrees, a total of about 8000 square degree, 1186 tiles, and about 50 tiles form a tile group, that is, each color block in the figure, a total of 24 tile groups.}
%  \label{fig7}
%\end{figure}

\subsection{Survey Scheduling Simulations}

Assuming 30 seconds exposure for one tile observation with 10 seconds telescope slew time, i.e., 40 seconds for each tile, it takes about 33 minutes for one tile group observation (i.e., a time block)\footnote{Simulations with accurate slew and readout time will be considered in future work.}. In terms of an 8,000\,deg$^2$ survey scheduling, a new tile group thus needs to be selected every 33 minutes, and the observing sequence of tiles in each group is not considered here.

To optimize the scheduling simulation, we applied the so-called Greedy Algorithm to arrange tile groups every night. This optimization algorithm is advantageous due to its simplicity, efficiency, and ease of implementation. The basic idea is to adopt the best or optimal choice in the current state at each step of selection, so that the final result is optimal. The optimization of the survey scheduling can be decomposed into sub-problems, specifically the optimization of each night or each individual observation. Before each observation, the tile group with the highest return metric is selected to obtain a local optimal solution. Once optimal observations are collected, the overall survey scheduling can be optimized. The return metrics for each tile group are generally composed by five return metrics:
\begin{enumerate}
  \item Sky background. Considering the moon position and moon phase influences on the sky background of each tile group.
  \item Altitude angle. The altitude angle of the target tile group is related to the airmass, as shown in Equation~\eqref{eq6}. A larger altitude angle corresponds to a smaller airmass, making it more favorable for observation.
  \item Angular distance. If the angular distance between current and next tile group is too large, which will negatively impact the observation efficiency due to a longer slew time.
  \item Historical observations. To ensure a relatively uniform total number of observations throughout the year in each tile group, the return metric should be reduced for areas with more historical observations. This will ensure that each tile group has an adequate amount of attention.
  \item Recent repeated observations. After establishing criteria based on the above four return metrics, it is possible that the return metric of a tile group is always higher than the other 23 tile groups in a specific time period, leading to the telescope to repeatedly observe the same tile group. Therefore, for tile groups that have been observed in previous two observing runs, the return metric is set to 0 in order to prevent the telescope from repeating observations.
\end{enumerate}

We note that the position coordinate of the central tile in each group is used for the calculation of the above return metrics. The values of the first four return metrics are linearly scaled between 0 and 10. For example, at a certain moment, the return metrics of the two tile groups with the darkest and brightest sky background are set to 10 and 0, while the return metrics of the other tile group are linearly scaled between 0 and 10 based on the background brightness. The same process applies to other return metrics. Therefore, a tile group with a return metric of 10 is the most sutiable group in terms of the imput observational conditions. Finally, the final optimization function is the maximum of the sum of the five return metrics for each tile group. When only a fraction of the tiles in a tile group meet observable conditions, the return metric of this tile group is calculated based on the central tile of this tile group. 

\begin{figure}[htb]
  \centering
  \includegraphics[width=1\columnwidth]{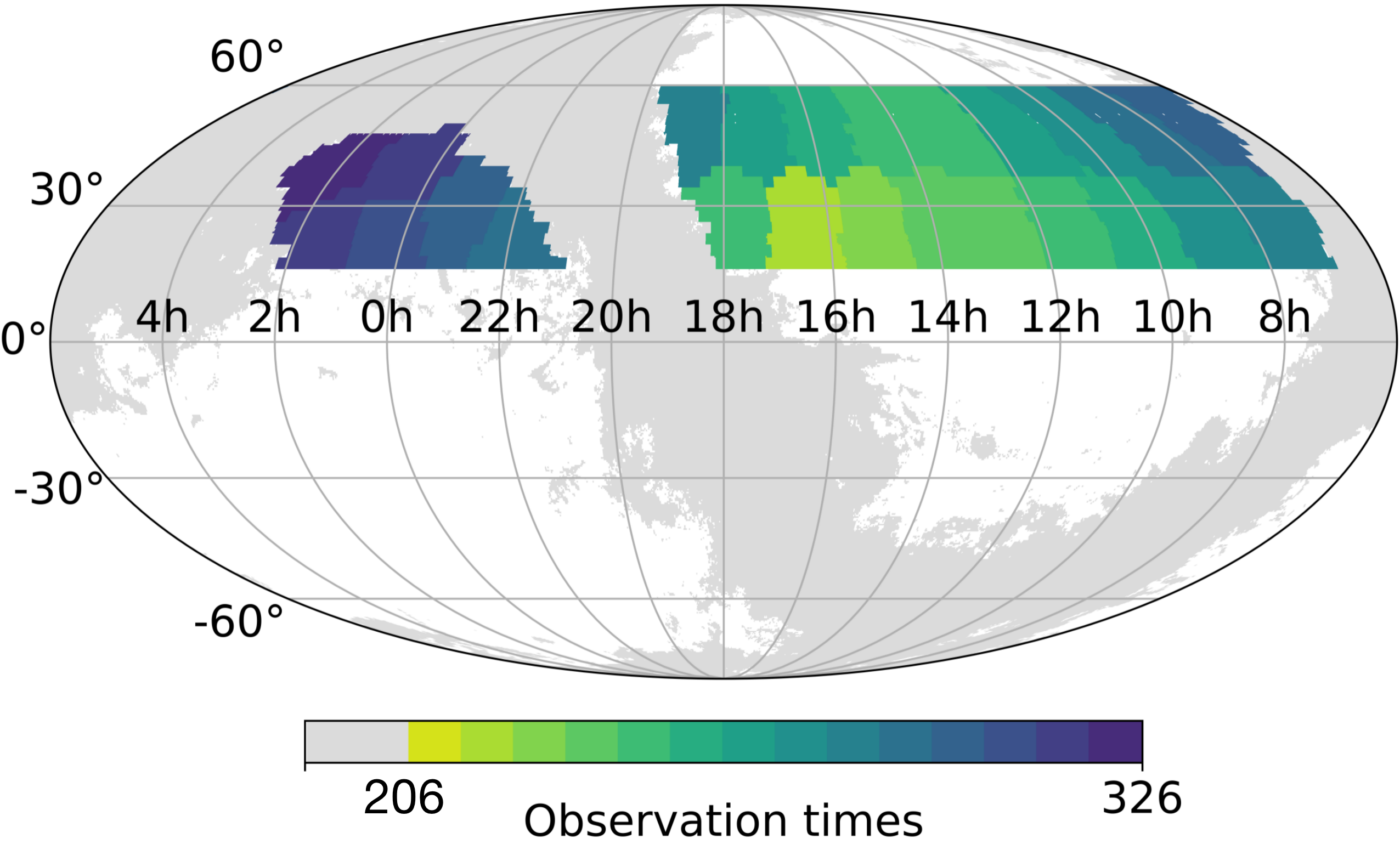}
  \caption{A tile group-based observation frequency distribution of a simulated 8,000 ${\rm deg}^2$ annual survey without considering weather and equipment downtime. Each tile group has about 260 observations throughout the year on average.}
  \label{fig8}
\end{figure}

Scheduling simulation results of an annual 8,000\,deg$^2$ survey are shown in Figure~\ref {fig8}. The simulation did not consider weather conditions or equipment downtime. The biggest number of observations in a year for a tile group is 326, while the smallest number is 206. Each tile group has an average of approximately 260 observations and a 130-minute exposure time in a year. For the future work, we will consider weather factors, filter switching, more detailed sleep time calculation,dithering pattern and equipment downtime factors, as well as the realtime clouds distribution at the Lenghu site.

\section{Dithering Pattern}
\label{sect:4}

A dedicated dithering pattern plays an important role in generating a high imaging uniformity of the final survey outputs. This technique involves slightly shifting the telescope's pointing between exposures to fill in gaps among CCDs. The layout diagram of the nine telescope focal plane arrays based on the division of the sky for WFST is shown in Figure~\ref{fig5.2}. The figure shows that there is an overlap between adjacent tiles, accounting for approximately 1.194\% of the area of a single tile. In addition, there is no data output at four corners of the tile and gaps among CCDs, which account for approximately 14.648\% of the entire tile. In other words, area with data output is about 84.158\%. If a telescope follows a fixed tiling method shown in the previous chapter, it will reduce the uniformity and efficiency of the survey program. The Rubin/LSST team has investigated various dithering patterns, developed a methodology for quantitatively comparing these patterns, and explored their effects on survey depth \citep{awan2016testing}. However, given the differences in both camera parameters and scientific requirements between Rubin/LSST and WFST, we newly developed a dithering pattern optimization method based on the Simulated Anneal algorithm \citep{kirkpatrick1983optimization}.

\subsection{Dithering pattern optimization}

According to the basic sky division mode of WFST, the minimum unit of division is a rectangle with dimensions of 2.65\,degrees and 2.6\,degrees for length and width, respectively. In order to optimize the dithering pattern, the lower-left corner of a tile is established as the coordinate origin, and a Cartesian coordinate system is set up with the X and Y axes along the lower and left edges of the tile, respectively. The length and width of the tile are then normalized to value 100, as shown in the central tile of Figure~\ref{fig5.2}. Consequently, without any dithering, the central coordinate of the tile are (50, 50). Once the dithering is conducted, the central coordinate of a tile will be changed accordingly. The center coordinate of a tile thus can be used to uniquely determine a dithering pattern. For example, in order to observe a certain sky area six times (with total five dithering), six pairs of (X, Y) coordinates are needed to express a specific dithering pattern.

In addition, there is a need for an indicator to evaluate the effectiveness of this dithering pattern, called the uniformity metric. The uniformity metric can be seen as the ratio of the pixel area with the expected number of visits within the tile to the total area, and the expected number of visits varies depending on the number of dithering observations. For example, in the case of 5-dithering observations, expected numbers of visits are set to five or six. In the case of 7-dithering observations, the expected numbers of visits are set to six, seven, and eight. Without any dithering pattern, the uniformity metric of this dithering pattern is 0.84158, because 84.158\% of the pixel areas within the tile reach the expected number of visits.

In general, there is a mathematical model of the dithering pattern, where the input of this mathematical model is the center coordinate of the tile observed $N$ times, and the output is the uniformity metric. With this model, the simulated annealing algorithm can be used to find the optimal dithering pattern with the highest uniformity metric for $N$ observations, i.e., the optimal dithering pattern. In the process of solving this mathematical model, the Python software package scikit-opt\footnote{https://scikit-opt.github.io/scikit-opt/} was used, which integrates many optimization algorithms, including simulated annealing algorithm.

\subsection{Results}

Once a dithering number can give a total blank area equals to the area of a tile, the number is considered as the least number that can give a relatively optimal uniformity. For example, if the area without data output of WFST is a regular rectangle and can perfectly fill the area of a tile, seven times ($1/0.14648=6.827$ rounded to 7) dithering observations should make the final image data relatively uniform. However, the irregular shape of blank area of WFST requires an additional dithering observation to make the area without data output as redundant as possible to fill the tile. Therefore, observing eight times should give a promising uniformity in principle.

\begin{figure*}[htb]
  \centering
  \includegraphics[width=\textwidth, angle=0]{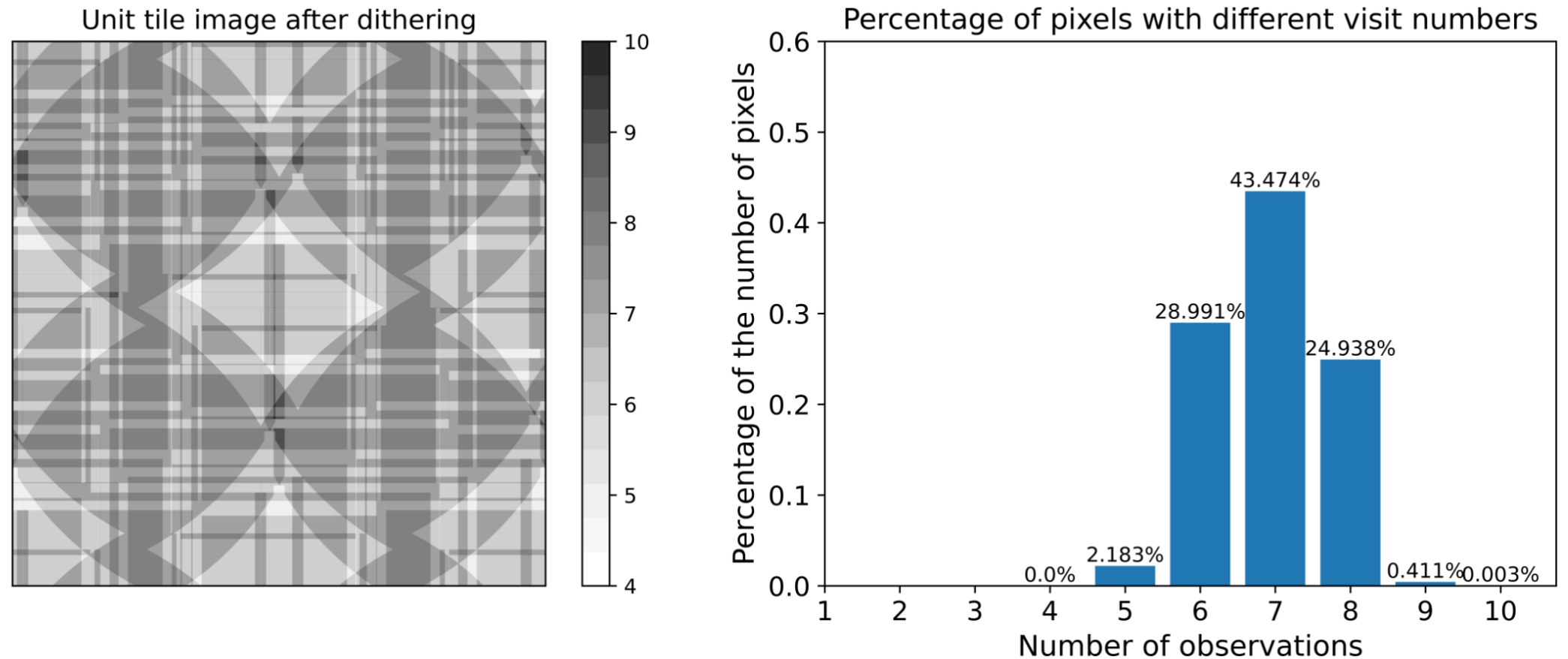}
  \caption{The optimized dithering pattern with eight observations, and the center coordinates of the eight observations are: (50, 50), (53.68, 92.76), (3.41, 98.93), (1.72, 43.31), (99.99, 53.28), (52.09, 40.36), (96.51, 96.03), and (48.07, 2.15), respectively. Left panel shows a footprint of eight pointings suggested by the optimized dithering pattern. Right panel shows the proportion of pixels in different visit numbers of a unit tile under the optimized dithering pattern.}
  \label{fig9}
\end{figure*}

By assuming eight times observations, we applied simulated annealing algorithm to calculate the optimized dithering pattern. The uniformity metric is set as the ratio of the area between six to eight observations to the area of the entire tile. The final dithering pattern is given in Figure~\ref{fig9}. The left figure illustrates the number of visits in each region after implementing the dithering pattern in a tile. The histogram on the right shows the proportion of 28.991\%, 43.474\%, and 24.938\% areas for six, seven, and eight times observations, respectively. The uniformity metric reaches 0.974, with almost no observation frequency below five times or above eight times, and the final output image is relatively uniform.

\section{Summary}
\label{sect:5}

We have performed a series of explorations on the observation capabilities of WFST before its official operation, and investigate the tiling method and dithering pattern optimization for the upcoming WFST surveys. Our main results are summarized as follows:
\begin{enumerate}
  \item The observing conditions of WFST were modeled, mainly considering four factors: site, meteorology, slew time, and sky background. If certain observing conditions are set, yearly observable time of each sky area in each band can be obtained. The average yearly observable time with airmass smaller than 1.5 in $u,g,r,i$ and $z$ is about 122, 167, 297, 355, and 355 hours, respectively.
  \item According to the effective coverage of the focal plane layout of WFST, the sky is divided into tiles. We further separated 8,000\,deg$^2$ sky area (about 1186 tiles) into 24 tile groups. The total observation time for a tile group is about 33 minutes. By applying the Greedy Algorithm with five return metrics: sky brightness, angular distance, altitude angle, historical observations, and recent repeated observations, the survey scheduling can be optimized on nightly. As a result, each tile can have an average of 260 observations per year without considering weather-out time and equipment downtime.
  \item When the telescope repeatedly observes the same sky area, a dedicated dithering pattern yields a high uniformity of the final survey outputs. Due to the focal plane layout of WFST, about 14.648\% of the survey area does not have data output when dithering is not performed. However, by applying a simulated annealing algorithm to optimize a dithering pattern of eight visits, we are able to achieve a 0\% sky areas without data output. Additionally, the proportion of sky areas observed less than six times is reduced to 2.183\%.
\end{enumerate}

We note that the current dithering pattern is limited to mid and low declination sky areas. The method of sky division introduced here can lead to non-negligible overlaps between adjacent tiles at high declination. Consequently, the dithering pattern for mid and low declination tiles may not be applicable for high declination tiles and requires future improvement. Although the Greedy Algorithm used in the current survey scheduling is efficient and easy to implement, it may not necessarily provide the global optimal solution. In the future, we plan to investigate other optimization algorithms and consider more delicate simulation to further improve the efficiency of survey scheduling.

\normalem
\begin{acknowledgements}
This work is supported by the National Science Foundation of China (12233005, 12073078, 12173088), the science research grants from the China Manned Space Project with NO. CMS-CSST-2021-A02, CMS-CSST-2021-A04 and CMS-CSST-2021-A07, and grants from the Cyrus Chun Ying Tang Foundations.

\end{acknowledgements}

% \onecolumn

\bibliographystyle{raa}
\bibliography{bibtex}

\end{document}